\documentclass{emulateapj}
\usepackage{graphicx}

\newcommand{\epsfig}[1]{\resizebox{\hsize}{!}{\includegraphics{#1}}}
\def\lsim{\mathrel{\rlap{\lower4pt\hbox{\hskip1pt$\sim$}}
    \raise1pt\hbox{$<$}}}          
\def\gsim{\mathrel{\rlap{\lower4pt\hbox{\hskip1pt$\sim$}}
    \raise1pt\hbox{$>$}}}          

\shorttitle{Munteanu et al.}
\shortauthors{Limit-cycle behavior in convective one-zone models}

\begin{document}

\title{Limit-cycle behavior in one-zone convective models}

\author{A.  Munteanu\altaffilmark{1},  G.  Bono\altaffilmark{2}, J.
Jos\'e\altaffilmark{3,4}, E.   Garc\'\i a--Berro\altaffilmark{1,3} and
R.F. Stellingwerf\altaffilmark{5}}

\altaffiltext{1}{Departament  de   F\'\i  sica  Aplicada,  Universitat
Polit\`ecnica de  Catalunya, Av.  del  Canal Ol\'\i mpic,  s/n, 08860,
Castelldefels (Barcelona) Spain; andreea@fa.upc.es; garcia@fa.upc.es}

\altaffiltext{2}{Osservatorio  Astronomico di  Roma, Via  Frascati 33,
00040 Monte Porzio Catone, Italy; bono@mporzio.astro.it}

\altaffiltext{3}{Institut  d'Estudis  Espacials  de Catalunya  (IEEC),
Ed. Nexus-201, c/ Gran Capit\`a 2-4, 08034, Barcelona, Spain}

\altaffiltext{4}{Departament  de  F\'\i  sica  i  Enginyeria  Nuclear,
Universitat  Polit\`ecnica de  Catalunya,  Av.  V\'{\i}ctor  Balaguer,
s/n,    08800   Vilanova   i    la   Geltr\'u    (Barcelona),   Spain;
jordi.jose@upc.es}

\altaffiltext{5}{Stellingwerf Consulting, 2229 Loma Linda, Los Alamos,
NM 87544; rfs@stellingwerf.com}

\received{April 29, 2004}
\accepted{March 21, 2005}

\begin{abstract} We present the results  of a detailed set of one-zone
models that account for  the coupling between pulsation and convection
following the original prescriptions  of \cite{S86}.  Motivated by the
arbitrary nature  of the input parameters adopted  in this theoretical
framework,  we  computed several  sequences  of  models  that cover  a
substantial fraction  of the parameter space and  a longer integration
time. We  found that our models  show the same  behavior as nonlinear,
hydrodynamical models,  that is, they approach  either the limit-cycle
stability  (pulsational instability) or  the fixed  point (pulsational
stability),  or present vibrational instability.   In agreement
with  \cite{S86},  we  find  that  convection is  the  main  quenching
mechanism   for  pulsational   models  located   across   the  Cepheid
instability  strip.   Moreover,  our  one-zone models  can  mimic  the
pulsational behavior of both  fundamental and first overtone Cepheids.
We  also  included a  turbulent  pressure  term  and found  that  this
physical   mechanism   plays  a   crucial   role   in  the   pulsation
characteristics of  the models  by removing the  sharp discontinuities
along the light  and the velocity curves showed by  models that do not
account  for turbulent  pressure.  Finally,  we investigated  the 
vibrational  and the pulsational  stability of  completely convective
models.  We consider as the most important finding of the present work
the  identification of a  well-defined region  in the  parameter space
where they approach limit-cycle stability.  The inclusion of turbulent
pressure  widens  this region,  thus  supporting original  suggestions
based on  both linear  and nonlinear models  of Long  Period Variables
(LPVs).  Several numerical experiments performed by adopting different
values of the  adiabatic exponent and of the  shell thickness indicate
that the coupling between pulsation  and convection is the key driving
mechanism  for  LPVs,  a   finding  supported  by  recent  theoretical
predictions.
\end{abstract}

\keywords{stars: oscillations, variables: Cepheids, Miras}

\section{Introduction}

Variable  stars are crucial  astrophysical objects  since they  can be
used  as tracers  of stellar  populations in  the framework  of galaxy
evolution  (\citealt{DEA02,MEA03}).  Moreover, the  comparison between
the observables  predicted by pulsational models  and the observations
themselves supply  an independent  estimate of the  stellar parameters
(\citealt{BEA01,KEA01}).  This implies that stellar pulsations provide
a  way  to  probe  regions   of  the  star  that  would  be  otherwise
unaccessible  to  direct   observations.   Therefore,  variable  stars
provide the unique opportunity  to investigate the plausibility of the
physical  assumptions  adopted  to  construct  both  evolutionary  and
pulsational models (\citealt{BCM02,KW02}).

Nonadiabatic  stellar pulsations are  complicated phenomena,  even for
small amplitudes  and purely radial oscillations, to  which the linear
theory  is applicable.   For large  amplitudes, the  nonlinear effects
become important and the  simplifying assumptions must be treated with
caution. During the last  decades the work on  nonlinear stellar
pulsations  has been  developed  along three  main approaches:  simple
one-zone   models,   the   formalism   of  amplitude   equations   and
hydrodynamical models. Although the  third approach provides the most
detailed and  accurate physical  description of the  outermost stellar
layers  during  the  pulsation cycle,  it  is  also  true that  it  is
difficult to figure  out whether the intrinsic features  of the models
are either a direct consequence of the adopted physical assumptions or
caused by the numerical methods  and the spatial resolution adopted to
discretize the  stellar structure into  a series of  concentric shells
--- see, e.g., \cite{PEA03} and references therein.  It is within
this       framework      where      simple       one-zone      models
(\citealt{B66,STT89,UX93,IFH92,T01}),   and  the   amplitude  equation
formalism  (\citealt{DK84,BK86,IT91,S93}),  play   a  key  role.   The
advantages and drawbacks of  these approaches have already been widely
discussed in the literature  --- see, e.g., \cite{B98}, and references
therein, and  \cite{T01}).  In  particular, one-zone models  have been
introduced  with  the  sole  purpose  of clarifying  the  analysis  by
eliminating possible subtle  uncertainties introduced by the stability
of  numerical  algorithms  and   by  spatial  resolution.   This  was
accomplished  by  considering  the  stellar  envelope  as  a  one-zone
structure: a single,  relatively thin, spherical mass-shell concentric
with   the  stellar   center.   These   models  have   provided  clear
understanding of  the destabilization  mechanisms and of  the possible
consequences  of  couplings  as  well  as  feedbacks  between  several
phenomena  associated  to  stellar  variability  ---  see  \cite{B66},
\cite{UW68}, \cite{STT89}, \cite{UX93}, \cite{IFH92}, and \cite{T01}.

The  simplest one-zone model  that accounts  for the  coupling between
pulsation and  convection was suggested by  \cite{S86}.  Following the
convective  scheme developed  by  \cite{S82a}, he  derived a  one-zone
pulsational model that includes  a time-dependent convective term.  He
computed a set of models  associated to the Cepheid instability strip,
integrating them only for a few dynamical time scales.  The results of
this work  support the  generally accepted view  that convection  is a
damping mechanism, in particular for  models located close to the cool
edge of  the instability  strip.  Although the  pulsationally unstable
region as well  as velocity curves were in  qualitative agreement with
empirical data,  he also found that fully  convective models underwent
vibrational instability for certain values of the parameters.  A
similar  approach   was  also  adopted  by   \cite{UX90,UX93}  and  by
\cite{S93}, but  they did  not discuss in  detail the  pulsational and
dynamical behavior of their models.

In this  work, we present new  results and possible  extensions of the
one-zone model developed  by \cite{S86}, investigating the limit-cycle
behavior (pulsational  instability) and by accounting for  the role of
the turbulent pressure  and of the thickness of  the convective layer.
The paper  is organized as  follows. In \S  2 we present  the one-zone
convective model.   We place special emphasis on  the adopted physical
and  numerical assumptions.   In \S  3 we  thoroughly  investigate the
limit-cycle behavior  of the model  suggested by \cite{S86},  an issue
marginally addressed in the original  paper.  In this section, we also
focus our attention  on the dependence of pulsation  properties on the
shell thickness (\S3.1)  as well as on the  turbulent pressure (\S3.2)
and their  role in  determining the morphology  of light  and velocity
curves.  In \S  4 we discuss the approach  to limit-cycle stability of
completely convective  models and briefly  describe possible empirical
similarities with LPVs.  Finally, in \S6 we summarize our main results
and briefly outline future perspectives.  In appendix A, we discuss in
more detail the physical  assumptions adopted to include the turbulent
pressure in the one-zone model are described.


\section{The one-zone convective model}

In the  classical theoretical  framework of one-zone  models, variable
stars have an equilibrium radius  $R_{\rm 0}$ and an extended shell or
envelope of  variable radius $R$  on top of  a compact core  of radius
$R_{\rm  c}$.   As  detailed  in  \cite{S86}, by  accounting  for  the
equation  of  motion,  the   energy  equation,  and  the  equation  of
convective  transport, the  resultant dynamical  system  describes the
evolution of the convective upper layer and it has the following form:

\begin{eqnarray}       &&\frac{d^2X}{d\tau^2}=HX^{-q}-X^{-2}\nonumber\\
&&\frac{dH}{d\tau}=\zeta       X^{2d}(1-\gamma_{\rm       r}       X^b
H^{s+4}-\gamma_{\rm    c}X^{-c}    U_{\rm    c}^3)\label{eq:dsystem}\\
&&\frac{d  U_{\rm c}}{d\tau}=\zeta_{\rm  c}  (X^{-d}H^{1/2}-U_{\rm c})
\nonumber,
\end{eqnarray}

\noindent  where $X$  and  $H$  are the  radius  and the  nonadiabatic
pressure normalized to their  equilibrium values, while $U_{\rm c}$ is
related to  the convective  velocity and it  is defined as  $U_{\rm c}
\equiv U'/U_{{\rm  ml}_0}$, where $U_{{\rm ml}_0}$  is the equilibrium
mixing-length  convective  velocity.   Moreover,  the  time  variable,
$\tau$,  is normalized  to the  dynamical (free-fall)  time  scale. As
defined in \cite{S86}, the other input parameters of the model are: $q
\equiv m  \Gamma_1-2$, $d\equiv m(\Gamma_1-2)/2$, and  $c \equiv m-2$,
where $\Gamma_1$  is the adiabatic  exponent and $m$ is  the so-called
form factor.  In the  limit of small oscillations (\citealt{S72,S86}),
the form factor is defined as:
                                           
\begin{equation}                                           m=\lim_{X\to
1}\frac{\log~[(X^3-\eta^3)/(1-\eta^3)]}{\log(X)}=
\frac{3}{1-\eta^3},\label{eq:m0}
\end{equation}

\noindent with $\eta \equiv R_{\rm  c}/ R_{\rm 0}$.  At the same time,
$b$ is  defined as  $b\equiv 4+m[n-(s+4)(\Gamma_1-1)]$, where  $s$ and
$n$  are the  temperature and  the density  exponents in  the Kramers'
opacity law --- for more details see \cite{B66}.

The main control parameters of the model are $\zeta$, $\zeta _{\rm c}$
and  $\gamma_{\rm c}$,  which  are  defined as  follows:  {\sl i)  the
nonadiabatic  parameter}  ---  $\zeta$   ---  the  ratio  between  the
dynamical  and  the  thermal  time  scale;  {\sl  ii)  the  convective
efficiency} ---  $\zeta_{\rm c}$ ---  the ratio between  the dynamical
time and the convective time scale; {\sl iii) the convective/radiative
splitting}  --- $\gamma_{\rm  c}$ ---  the ratio  between  the initial
convective  luminosity  and the  total  initial  luminosity, that  is,
$\gamma_{\rm  c}   \equiv  L_{{\rm  c}_0}/L_0$,   which  implies  that
$\gamma_{\rm r} = 1-\gamma_{\rm c}$.

The reader interested  in a  detailed discussion  concerning the
convective time scale in the context of one-zone models is referred to
\S  4 of  \cite{S86}.   More quantitative  predictions concerning  the
different  timescales  connected  with  turbulent  energy  across  the
envelope  of  variable  stars   have  been  provided  by  \cite{S82a},
\cite{BMS99} and by \cite{FBK00}.

The values $n=1$  and $s=3$ for the Kramers'  opacity are the standard
ones, while the values  for $\Gamma_1$, $m$, $\gamma_{\rm c}$, $\zeta$
and  $\zeta_{\rm c}$  must  be  treated with  caution.   The value  of
$\Gamma_1=1.1$  adopted  in \cite{S86}  is  a  typical  value for  the
$\gamma$-mechanism  operating   in  the  partial   ionization  regions
(\citealt{C80}).   The  choice  of  a  form factor  $m$  equal  to  10
($\eta=0.888$) ---  considered typical  of Cepheids in  \cite{S86} ---
together with  the values chosen  for the other parameters  assure the
turbulent  stability  and  the   secular  stability  for  $\zeta$  and
$\zeta_{\rm  c}$ in  the range  $[0,10]$  and $\gamma_{\rm  c} \le  1$
(\citealt{S86}).   The  values of  these  parameters were  arbitrarily
chosen  by \cite{S86}  and lead  to the  existence of  a pulsationally
unstable region  (``strip'') for values of $\gamma_{\rm  c} \le 0.45$.
From a  mathematical point  of view,  we refer to  such a  behavior as
having a limit  cycle born through a Hopf  bifurcation.  The stability
analysis in \cite{S86} was hampered by the fact that individual models
were integrated only for a  small number of dynamical time scales.  In
particular, our calculations show that for these cases, no limit-cycle
stability exists.   Note also that \cite{S86} analyzed  only the cases
for $\zeta$ and $\zeta_{\rm c}\le  4$, and that an exhaustive study of
the  dependence of the  limit-cycle behavior  on the  input parameters
remained to be performed.

\begin{figure}[t]
\epsfig{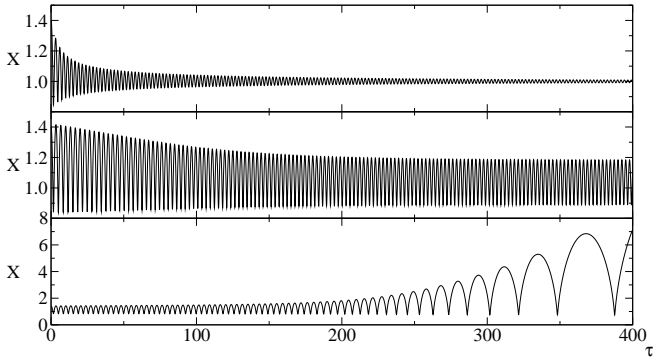}
\caption{Examples  of dynamical behavior  for $\gamma_{\rm
c}=0.4$.    Radius   time   series   for   stable   ---   $(\zeta_{\rm
c},\zeta)=(3,4)$  ---   {\sl  (top)},  limit-cycle   ---  $(\zeta_{\rm
c},\zeta)=(1.5,1.5)$  ---  {\sl (middle)}  and  unstable behavior  ---
$(\zeta_{\rm c},\zeta)=(0.5,0.5)$ --- {\sl (bottom)}. \label{f01}}
\end{figure}

\begin{figure*}[t]
\epsfig{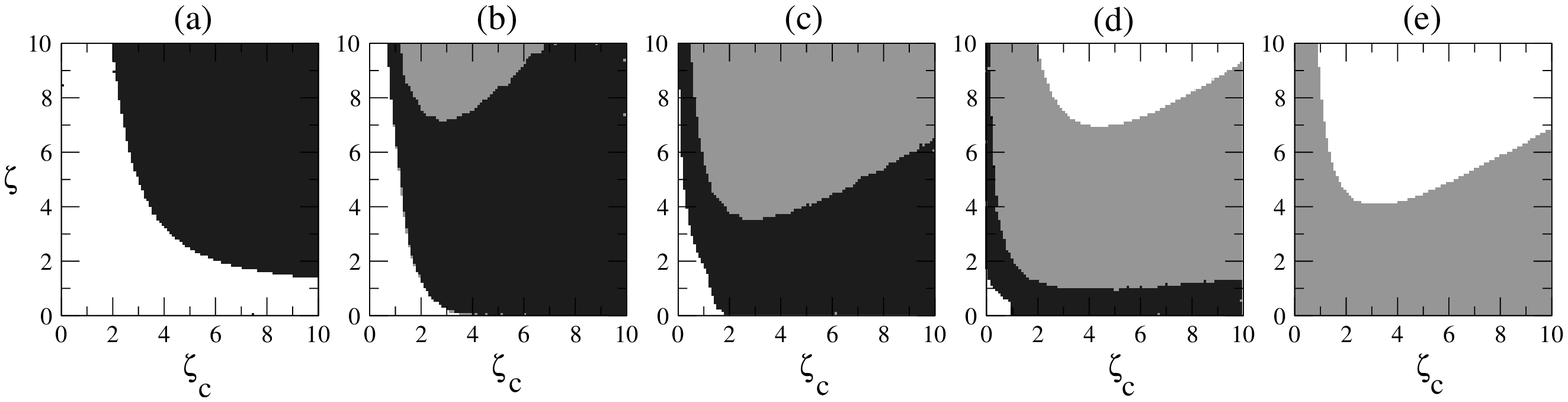}
\caption{The  ($\zeta,\zeta_{\rm  c}$)-plane  for  several
values   of  $\gamma_{\rm   c}$:  $\gamma_{\rm   c}=0.1$   {\sl  (a)};
$\gamma_{\rm  c}=0.2$  {\sl  (b)};  $\gamma_{\rm  c}=0.3$  {\sl  (c)};
$\gamma_{\rm  c}=0.4$  {\sl  (d)};  $\gamma_{\rm  c}=0.5$  {\sl  (e)},
representing  pulsationally   unstable  {\sl  (black)},  pulsationally
stable  {\sl (grey)}  and   vibrationally unstable  behavior {\sl
(white)}}
\label{f02}
\end{figure*}

\begin{figure*}[t]
\epsfig{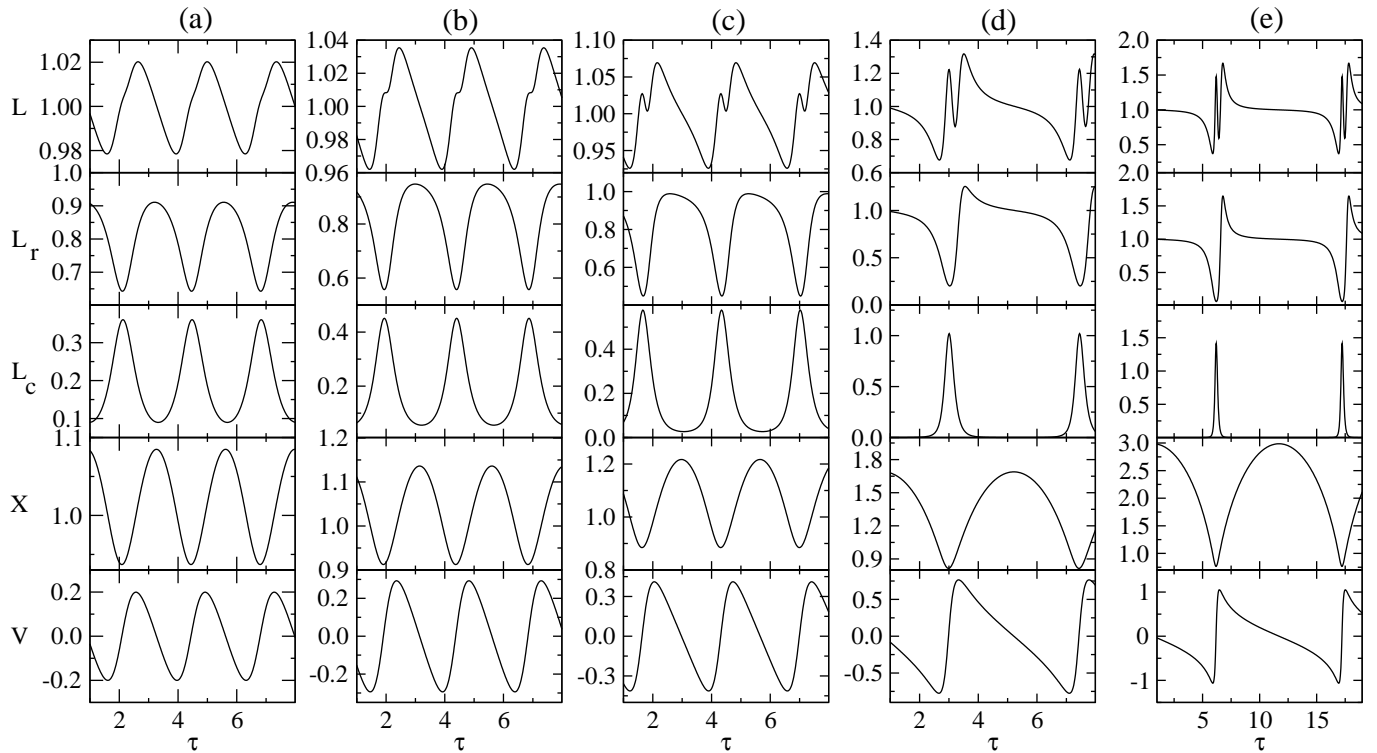}
\caption{Limit-cycle  characteristics. From top  to bottom
the panels show: the  total luminosity, $L$; the radiative luminosity,
$L_{\rm  r}   \equiv  {L_{\star}}_{\rm  r}/L_0$   and  the  convective
luminosity,  $L_{\rm  c}  \equiv  {L_{\star}}_{\rm  c}/L_0$;  and  the
dimensionless radius,  $X$ and velocity, $V$  variations. The temporal
axis is represented as $\tau -\tau_{\rm lc}$, where $\tau_{\rm lc}$ is
the  time when  the  limit  cycle has  been  reached.  The  parameters
adopted   to  construct   the  individual   cases  were   chosen  from
Fig.~\ref{f02}{\sl b}: {\sl  (a)} Small amplitude: $(\zeta, \zeta_{\rm
c})=(6.5,3)$;   {\sl   (b)}    Bump   Cepheid:   $(\zeta,   \zeta_{\rm
c})=(5.5,3)$;  {\sl  (c)}  Double-peak  Cepheid:  $(\zeta,  \zeta_{\rm
c})=(4,3)$;  {\sl (d)} Steep  bump: $(\zeta,  \zeta_{\rm c})=(1.5,3)$;
{\sl (e)} Pulse-like: $(\zeta, \zeta_{\rm c})=(0.7,3)$.\label{f03}}
\end{figure*}

In  the current  investigation, we  have first  undertaken  a thorough
study  of the  dynamics  in  the $(\zeta,  \zeta_{\rm  c})$ plane  for
$\zeta$ and  $\zeta_{\rm c} \leq 10$  and for different  values of the
convective-radiative   splitting,  $\gamma_{\rm  c}$,   following  the
assumption  of radiative-dominated  energy transport  ($\gamma_{\rm c}
\leq 0.5$).  We  have used the same form factor  ($m=10$) and the same
initial  condition  $(X_0,V_0,H_0,U_{{\rm  c}_0})=  (1.4,0.0,1.0,0.7)$
adopted by  \cite{S86}.  We  have furthermore assumed  that a  case is
pulsationally  unstable (limit-cycle  behavior)  when two  consecutive
maxima  in  radius  variations  differ  by less  than  $10^{-7}$,  and
pulsationally   stable  (damping   oscillations)  when   the  solution
asymptotically approaches the fixed point $(\bar{X}, \bar{V}, \bar{H},
\bar{U_{\rm  c}}) =  (1,0,1,1)$ and  its distance  to the  fixed point
becomes smaller  than $10^{-8}$.  The solution has  been considered to
be  vibrationally unstable when  the rapidly  increasing dimensionless
radius  amplitude  became  larger   than  15.   These  thresholds  are
reasonable values.  Moreover, we have also verified that the resultant
qualitative dynamics  does not depend  on the choice of  these values.
Of course, one could use as well the asymptotic perturbation theory to
compute  the properties  of the  limit cycle.   Nevertheless,  such an
approach  is beyond  the  aims  of the  current  investigation and  we
consider it as  a future line of work.  For  a better understanding of
our  results,  we illustrate  in  Fig.~\ref{f01}  the  three types  of
dynamical behavior we have identified.  The top panel shows an example
of pulsational stability, i.e.,  the initial perturbation decays.  The
bottom  panel  shows  a  vibrationally  unstable  model  (the  initial
perturbation grows), while the middle panel shows the time series of a
case   that   approaches    a   limit-cycle   stability   (pulsational
instability).  Note  that for small  amplitudes ($\tau \leq  100$) the
case plotted in the bottom panel  seems to approach a limit cycle, but
for $\tau  > 200$  it becomes clear  that the amplitudes  are steadily
increasing.  This  finding further strengthens the need  for long time
integrations  to  assess the  type  of  stability  for the  individual
one-zone  models.  As  a final  comment on  the  numerical assumptions
considered  in the  present  work,  it is  worth  mentioning that  the
dynamical  behavior  of  the  system  is independent  of  the  initial
condition, provided that  it belongs to the neighborhood  of the fixed
point  $(\bar{X},  \bar{V},  \bar{H},  \bar{U_{\rm  c}})=  (1,0,1,1)$.
Therefore,   we   did   not   investigate  the   dependence   of   the
characteristics of the time-series on initial conditions.


\begin{figure*}[t]
\epsfig{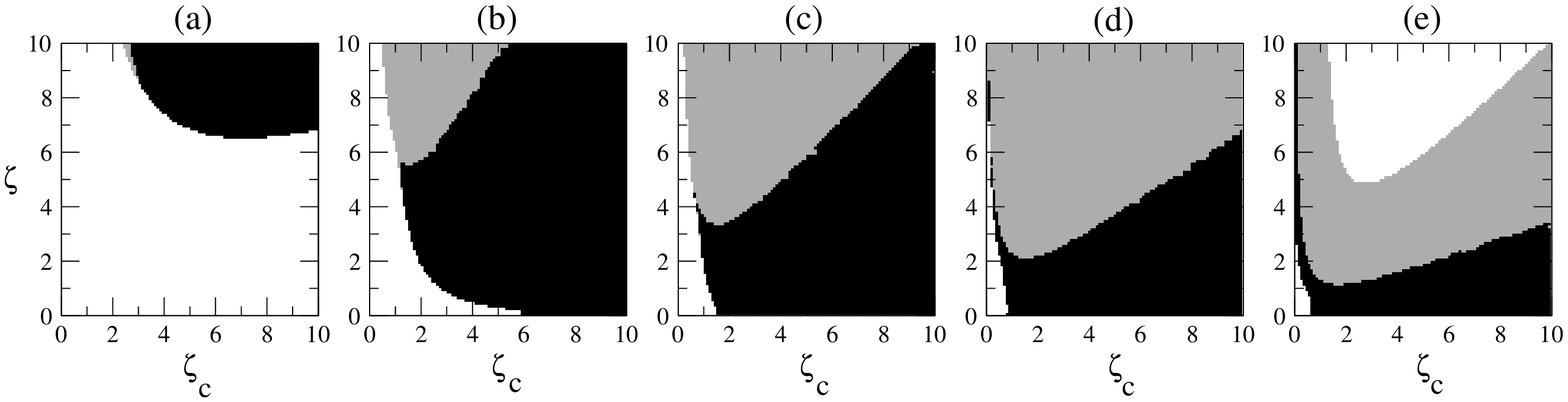}
\caption{The  ($\zeta,\zeta_{\rm  c}$)-plane  for  several
values of $\gamma_{\rm c}$.  Eq.~(\ref{eq:dsystem}) has been used with
$\eta=0.75$: $\gamma_{\rm c}=0.1$ {\sl (a)}; $\gamma_{\rm c}=0.2$ {\sl
(b)}; $\gamma_{\rm c}=0.3$ {\sl  (c)}; $\gamma_{\rm c}=0.4$ {\sl (d)};
$\gamma_{\rm  c}=0.5$ {\sl  (e)}, representing  pulsationally unstable
{\sl   (black)},   pulsationally   stable   {\sl  (grey)}   and  
vibrationally unstable behavior {\sl (white)}. \label{f04}}
\end{figure*}

\begin{figure*}[t]
\epsfig{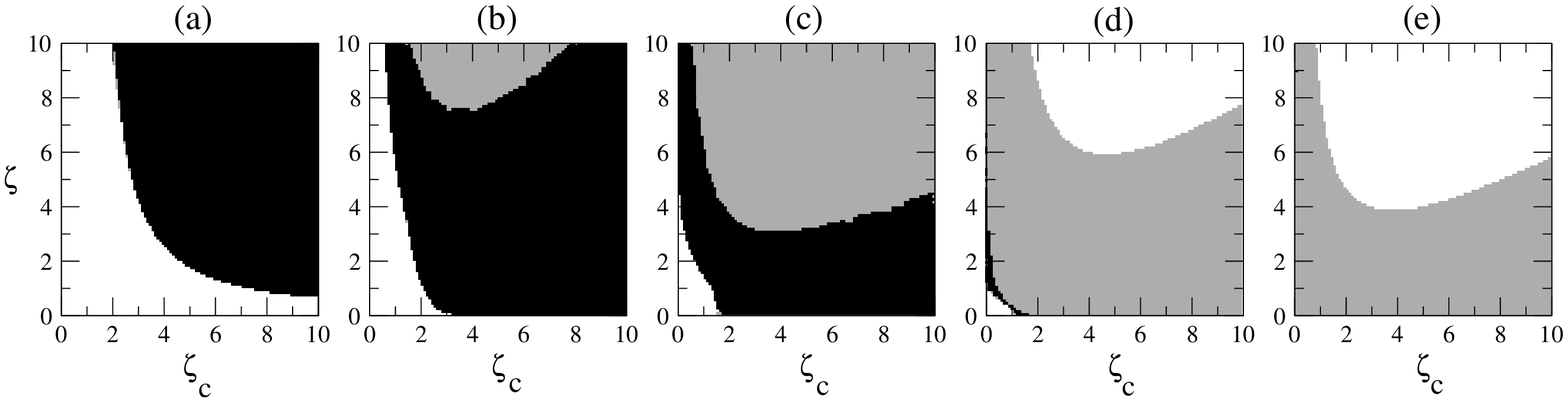}
\caption{The  ($\zeta,\zeta_{\rm  c}$)-plane  for  several
values of $\gamma_{\rm c}$.  Eq.~(\ref{eq:dsystem}) has been used with
$\eta=0.92$: $\gamma_{\rm c}=0.1$ {\sl (a)}; $\gamma_{\rm c}=0.2$ {\sl
(b)}; $\gamma_{\rm c}=0.3$ {\sl  (c)}; $\gamma_{\rm c}=0.4$ {\sl (d)};
$\gamma_{\rm  c}=0.5$ {\sl  (e)}, representing  pulsationally unstable
{\sl   (black)},   pulsationally   stable   {\sl  (grey)}   and  
vibrationally unstable behavior {\sl (white)}. \label{f05}}
\end{figure*}

\section{Limit-cycle characteristics}

One-zone models  can be  adopted to  perform only  a qualitative
comparison with pulsation properties of variable stars. Moreover, this
comparison is limited to  observables that one-zone models can account
for.   In particular,  the  shape  of light  and  velocity curves  ---
originally suggested  for RR  Lyrae stars by  \cite{SD86} ---  and the
occurrence of  first overtone pulsators  (\citealt{SGD87}).  The most
interesting  results disclosed  by  our simulations  are presented  in
Fig.~\ref{f02}{\sl a--e}, where the regions of damped oscillations are
shown in grey,  the regions where the radial  displacements approach a
limit cycle  are marked in  black, while vibrationally unstable
cases  are  shown  in  white.    A  glance  at  the  data  plotted  in
Fig.~\ref{f02} discloses that, with the exception of the case in which
$\gamma_{\rm  c}=0.1$,  an  increase  in  $\gamma_{\rm  c}$  causes  a
decrease in  the region of  pulsationally unstable cases and  that for
$\gamma_{\rm c}=0.5$  it vanishes. These findings  support the results
originally  obtained  by \cite{S86}  concerning  the  damping role  of
convection.  In addition to the radius $(X)$ and velocity $(V)$ curves
chosen to be illustrated in  \cite{S86}, we have added the light $(L)$
curve  and  in several  occasions  the  temperature $(T)$  variations,
estimated according to

\begin{eqnarray}                       L                      &\equiv&
\frac{L_{\star}}{L_0}=\frac{{L_{\star}}_{\rm    r}+   {L_{\star}}_{\rm
c}}{L_0}= \gamma_{\rm r} X^bH^{s+4}+ \gamma_{\rm c} X^{-c} U_{\rm c}^3
\\ \frac{T}{T_0} &=& X^{-2 d} H,
\end{eqnarray}

\noindent where  $L_{\star}$ denotes the stellar  luminosity while the
subscripts  ``r'' and  ``c''  refer to  the  radiative and  convective
component,   respectively,    as   they   result    from   \cite{S86}.
Fig.~\ref{f03}  shows the light  curve, $(L)$,  the radius  $(X)$, and
velocity  $(V)$ time  series for  selected cases  that  approach limit
cycle.  Note that to improve the visualization of data plotted in Fig.
~\ref{f03}, we subtracted from  the integration time, $\tau$, the time
interval  spent  by individual  models  to  approach  the limit  cycle
stability.  We  have chosen to illustrate cases  at fixed $\gamma_{\rm
c}$  and $\zeta_{\rm  c}$ values,  and $\zeta$  ranging from  $0.7$ to
$6.5$, being typical examples of the qualitative dynamics encountered.
The  time  series   plotted  in  this  figure  show   that  the  cases
characterized  by  parameters  located   close  to  the  edge  of  the
pulsational  stability region  (grey area  in  Fig.~\ref{f02}) present
small  amplitudes  and  sinusoidal  changes (panel  {\sl  a}).   These
features are quite similar  to the behavior that nonlinear, convective
models show close  to the blue edge of the  instability strip --- see,
e.g., \cite{BCM00}.  On the other hand, models that are located across
the  pulsationally unstable  region (panels  {\sl b,  c}) show  a bump
along  the rising branch  of the  luminosity curve.   Our calculations
reveal  that such  a  feature is  due  to the  sharp  increase in  the
efficiency of  the convective motions,  close to the phase  of minimum
radius.   This  finding  is  also found  in  nonlinear  hydrodynamical
models.

\begin{figure*}[t]
\epsfig{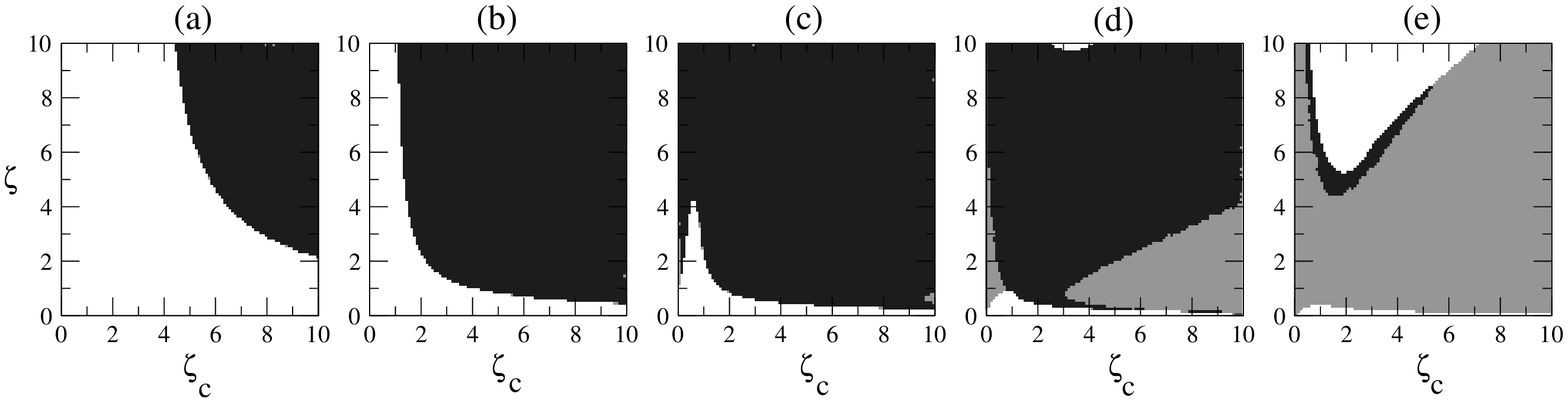}
\caption{The     ($\zeta,\zeta_{\rm     c}$)-plane     for
$\eta=0.888$ and several values of $\gamma_{\rm c}$ when the turbulent
pressure  is included:  $\gamma_{\rm c}=0.1$  {\sl  (a)}; $\gamma_{\rm
c}=0.2$  {\sl  (b)};  $\gamma_{\rm  c}=0.3$  {\sl  (c)};  $\gamma_{\rm
c}=0.4$  {\sl  (d)};  $\gamma_{\rm  c}=0.5$  {\sl  (e)},  representing
limit-cycle {\sl (black)}, stable  {\sl (grey)} and vibrationally
unstable behavior {\sl (white)}. \label{f06}}
\end{figure*}

\begin{figure*}[t]
\epsfig{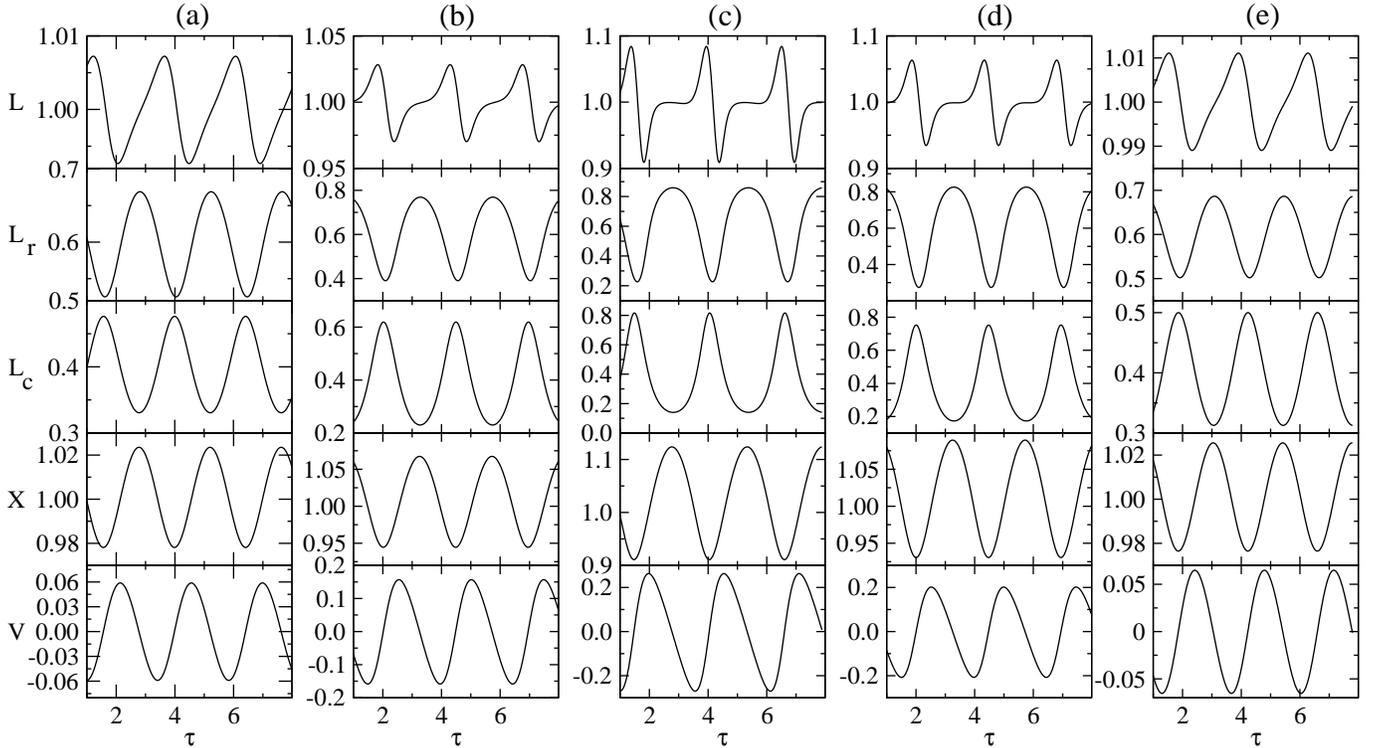}
\caption{Limit-cycle   characteristics   when  the   turbulent
pressure is introduced. From top  to bottom the panels show: the total
luminosity,   $L$;  the  radiative   luminosity,  $L_{\rm   r}  \equiv
{L_{\star}}_{\rm  r}/L_0$ and  the convective  luminosity,  $L_{\rm c}
\equiv {L_{\star}}_{\rm c}/L_0$; and the radius, $X$ and the velocity,
$V$ variations.  The temporal  axis is represented as $\tau -\tau_{\rm
lc}$, where $\tau_{\rm lc}$ is the time value when the limit cycle has
been  achieved.  The  parameters adopted  to construct  the individual
cases are:  $\eta=0.888$, $\gamma_{\rm c}=0.4$ and  {\sl (a)} $(\zeta,
\zeta_{\rm c})=(4,0.28)$; {\sl  (b)} $(\zeta, \zeta_{\rm c})=(4,0.4)$;
{\sl   (c)}  $(\zeta,  \zeta_{\rm   c})=(4,1)$;  {\sl   (d)}  $(\zeta,
\zeta_{\rm     c})=(4,4)$;    {\sl     (e)}     $(\zeta,    \zeta_{\rm
c})=(4,9)$. \label{f07}}
\end{figure*}

One can notice that for  cases located close to the vibrationally
unstable region,  the bump  becomes very narrow  (panel {\sl  d}) and
pulse-like (panel {\sl e}). This is a behavior that is not observed in
actual  Cepheids and  it  is not  supported  by nonlinear,  convective
models.  This indicates the limit  of the crude physical and numerical
approximations adopted to construct our models.  However, our analysis
suggests that the limit-cycle behavior of these one-zone models in the
pulsationally  unstable region  mimics the  behavior of  the Classical
Cepheids  instability  strip,  as  argued  in the  original  paper  of
\cite{S86}.  Finally, we note that the period of the oscillations
of the  cases plotted in  Fig. 3 is  not unity, even if  the dynamical
timescale was normalized to unity. This strong nonlinear effect on the
period was already  noted by \cite{S86}, and he  suggested that it was
due  to   a  correlation  with  the  pulsation   amplitude  (see  also
\citealt{SD86}).
 

\subsection{The shell thickness}

In this section,  we present the results concerning  the impact of the
shell thickness on the existence  of a limit-cycle behavior and on its
characteristics.    According   to   classical   physical   arguments,
pulsational  models only account  for the  envelope of  variable stars
(\citealt{C80}).   They include  a damping,  adiabatic region  that is
typically  located on  top of  the  nuclear burning  region (core),  a
transition region,  a driving, nonadiabatic region,  and the outermost
layers (surface).  The  driving region is the envelope  zone where key
elements (hydrogen, helium, metals)  are partially ionized and supply,
via  the  $\kappa$-   and/or  the  $\gamma$-mechanism,  the  pulsation
destabilization.   In  this theoretical  framework,  the  base of  the
envelope in nonlinear, hydrodynamical models is typically located at a
few percents of the total radius (\citealt{PEA03}).

As  far  as  the   one-zone-model  approach  is  concerned,  different
assumptions have  been adopted in  the literature. The  one-zone model
suggested by  \cite{IFH92} for Mira  variables accounts for  a driving
region  (piston  approximation) at  $X\approx  0.2-0.4$, a  transition
region through  which the pressure waves from  the interior propagate,
and a dissipation region that they call mantle. On the other hand, the
one-zone  model  of  \cite{S86}  for  variable stars  in  the  Cepheid
instability strip accounts for  a driving convective region located at
$X\approx  0.85$ on  top of  a damping,  adiabatic region  ({\sl rigid
core}). In  this theoretical framework,  the pulsation destabilization
is provided by  a driving agent ($\Gamma_1 <  4/3$). The two different
assumptions describe the same envelope region if the {\sl rigid core},
assumed by  \cite{S86} is  the outer boundary  of the  damping region.
Therefore,  the  {\sl shell  thickness},  $(1-\eta)$ in  dimensionless
formulation,  is the  radial extent  of the  region located  above the
boundary   between   the   damping   (adiabatic)   and   the   driving
(nonadiabatic) region.

To investigate  the dependence of  the dynamic behavior of  the system
given in Eq.~(\ref{eq:dsystem}) on the shell thickness, we have chosen
two values  of the shell thickness,  one smaller and  the other larger
than the value adopted in \cite{S86}: $\eta=0.75$ and $\eta=0.92$.  We
illustrate in  Fig.~\ref{f04} and Fig.~\ref{f05} the  behaviors in the
($\zeta,\zeta_{\rm c}$)-plane for $\gamma_{\rm  c} \leq 0.5$.  One can
notice that the  increase in the shell thickness  (decrease of $\eta$)
leads  to   the  existence   of  pulsational  instability   for  cases
characterized  by stronger convection  (higher values  of $\gamma_{\rm
c}$). This effect is expected, because a decrease in $\eta$ implies an
increase in the extent of the driving region.  Moreover, we have noted
that for  a fixed  value of $\gamma_{\rm  c}$ and different  values of
$\eta$, the  period distribution of  the cases that approach  a stable
limit cycle peaks at shorter periods as the shell thickness decreases.
This   behavior   is   also   expected,  since   detailed   nonlinear,
hydrodynamical  models  of  RR  Lyrae (\citealt{BS94})  and  classical
Cepheids (\citealt{BMS99}) suggest that  the regions located below the
nodal line supply  a small contribution to the  work integral of first
overtone pulsators.   This supports the suggestion  by \cite{SGD87} to
decrease the shell thickness of one-zone models to mimic the dynamical
behavior of overtone pulsators.


\subsection{The turbulent pressure}

Convection affects pulsation  through three factors, namely convective
energy  transfer  (thermodynamic  coupling), turbulent  pressure,  and
turbulent  viscosity  (dynamic  coupling).   The effect  of  turbulent
viscosity  is to  convert the  kinetic energy  of radial  motions into
thermal energy by means of  a turbulent cascade of smaller and smaller
turbulent eddies.  This  means that the turbulent viscosity  is a pure
damping factor (\citealt{S82a,XDC98a,YKB98}).  The role that turbulent
pressure   plays  in  driving   or  damping   the  pulsation   is  not
straightforward, since the contributions of gas and turbulent pressure
can not  be easily  separated. Linear (\citealt{YKB98})  and nonlinear
(\citealt{BMS99}) convective models of Classical Cepheids suggest that
the work  done by turbulent  pressure attains both  positive (driving)
and   negative  (damping)   values   in  different   regions  of   the
envelope. Linear  and nonlinear, convective LPV  models constructed by
\cite{OC86} and  by \cite{CO93} indicate that turbulent  pressure is a
driving mechanism,  whereas more recent  calculations by \cite{XDC98a}
support the evidence that it is a damping mechanism.

The turbulent pressure, $P_{\rm t}$,  was not included in the original
one-zone model by \cite{S86}, but it was shortly mentioned as being

\begin{equation} \frac{P_{\rm t}}{P_{\rm t_0}}=X^{-m}U_{\rm c}^2,
\label{eq:turbulence}
\end{equation}

\noindent where the zero  subscript denotes its equilibrium value.  In
this section, we  present the results obtained by  the introduction of
the  turbulent   pressure.   The  reader  interested   in  a  detailed
description  of  the  physical  assumptions  adopted  to  include  the
turbulent  pressure  term  in  Eq.~(\ref{eq:dsystem}) is  referred  to
Appendix A. The new dynamic system becomes now:

\begin{eqnarray}                  \frac{d^2X}{d\tau^2}&=&(1-\alpha_{\rm
p})X^{-q}h+\alpha_{\rm    p}X^{-c}    U_{\rm    c}^2-X^{-2}\nonumber\\
\frac{dH}{d\tau}&=&-m~\frac{\alpha_{\rm p}(\Gamma_3-1)}{1- \alpha_{\rm
p}}~X^{2d-1}~\frac{dX}{d\tau}~U_{\rm    c}^2~-\nonumber\\    &-&~\zeta
X^{2d}~(\gamma_{\rm  r}X^b  h^{s+4}+(1-\gamma_{\rm  r}) X^{-c}  U_{\rm
c}^3-1) \label{eq:dsystemtp}\\ \frac{d U_{\rm c}}{d\tau}&=& \zeta_{\rm
c}(X^{-d}H^{1/2}-U_{\rm c}) \nonumber,
\end{eqnarray}

We  have undertaken  a parametric  study in  order to  investigate the
influence of  the turbulent  pressure on the  overall dynamics  of the
system  and implicitly  on the  existence  of limit  cycles. For  this
purpose, we have chosen the same initial condition as in \S 2, that is
$(X_0, V_0, H_0, U_{{\rm c}_0})=(1.4,  0.0, 1.0, 0.7)$, which led to a
value  of $\alpha_{\rm p}  \approx 0.4$.   The results  concerning the
plane $(\zeta,\zeta_{\rm  c})$ are shown in  Fig.~\ref{f06}, using the
color  code from  Fig.~\ref{f02}.   The regions  characterized by  the
existence of limit cycles for $\gamma_{\rm c} < 0.5$ are more extended
than in the  case without turbulent pressure, as  expected because the
additional pressure adds to the driving mechanism.

We have also investigated the  light and velocity curves for the cases
characterized by limit-cycle behavior.  From the hydrodynamical models
as  well as  from plain  physical arguments  on the  existence  of the
Cepheid  instability  strip,  it  is  expected that  by  crossing  the
instability strip from the blue  to the red edge, the period increases
and the amplitude of the  oscillations quickly reaches its maximum and
then slowly decreases.  In order  to obtain a similar evolution of the
amplitude  from this one-zone  model, the  limit-cycle region  must be
bounded in  the $(\zeta,\zeta_{\rm c})$-plane  by pulsationally stable
regions both  to the left (low  values of $\zeta_{\rm c}$)  and to the
right (high values of $\zeta_{\rm c}$).  If bounded by a vibrationally
unstable region,  the oscillations which result  from using parameters
close to this  boundary are pulse-like and the  amplitude is large, as
in  Fig.~\ref{f03}{\sl  e}.  One  can  notice  that  for the  case  of
zero-turbulence  pressure, the  limit-cycle region  is bounded  to the
left  by  the  vibrationally  unstable  behavior and  thus  the  above
condition  is not satisfied.   However, with  the introduction  of the
turbulent  pressure, there  exists a  case for  which  the limit-cycle
region  is bounded in  the $(\zeta,\zeta_{\rm  c})$-plane both  to the
left  and   to  the  right   by  the  pulsationally   stable  behavior
(Fig.~\ref{f06}{\sl d}).  For this  case, we present in Fig.~\ref{f07}
the types of light and velocity curves obtained in the transition from
the hot (blue, low $\zeta_{\rm c}$) to the cool (red, high $\zeta_{\rm
c}$) edge, at constant $\zeta$.  One can notice that the amplitude has
a  peak  toward the  center  of the  limit-cycle  region  and then  it
decreases.   Moreover  and  even  more importantly,  data  plotted  in
Fig.~\ref{f07}  disclose   that  one-zone  models   that  account  for
turbulent pressure  do not  present, in this  region of  the parameter
space,  the spurious  secondary  peak along  the  rising branch.   The
morphology of light and velocity  curves plotted in this figure are in
qualitative  agreement   with  actual  classical   Cepheids  and  with
nonlinear, convective models.


\section{The red variables}

\begin{figure}[b]
\epsfig{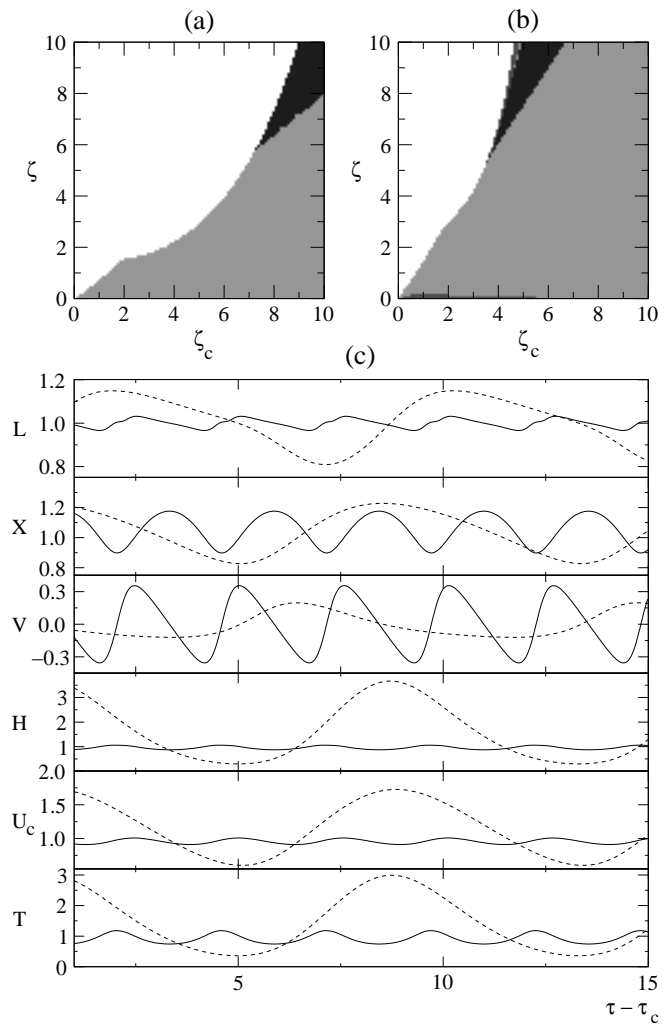}
\caption{The ($\zeta,\zeta_{\rm  c}$)-plane for the completely
convective  shell  illustrated in  the  spirit  of Fig.\ref{f02}  for:
$\gamma_{\rm  c}=1.0$,  $\eta=0.888$   and  {\sl  (a)}:  no  turbulent
pressure, and  {\sl (b)}: turbulent pressure included;  {\sl (c)} Time
series for the luminosity $L$, radius $X$, velocity $V$, pressure $H$,
convective velocity $U_{\rm c}$ and temperature $T$ for a Cepheid-like
(solid line: $\gamma_{\rm c}=0.2$, $\zeta_{\rm c}=1.2$, $\zeta=8$) and
a   LPV-like  variable  star   (dashed  line:   $\gamma_{\rm  c}=1.0$,
$\zeta_{\rm c}=9$,  $\zeta=7.5$), the  latter case was  extracted from
panel {\sl(a)}. \label{f08}}
\end{figure}

\cite{S86}  remarked a  paradox of  the present  one-zone  model: {\sl
although  the  convection tends  to  stabilize  the  pulsation in  the
majority of cases, the completely convective case shows instability in
every criterion}.  We have verified that indeed the present model with
fully   convective  ($\gamma_{\rm   c}=1$)  thick   shell   ($m<8$  or
$\eta<0.85$)  presents   vibrational  instability  for   any  $\zeta$,
$\zeta_{\rm c}  \leq 10$. However,  for a thin shell,  our simulations
revealed that  while no limit-cycle  regions exist for the  cases with
$0.5  \lsim \gamma_c  < 1.0$  and $\zeta,  \zeta_{\rm c}  \in [0,10]$,
limit-cycle  models exist  for the  case of  fully  convective shells.
Fig.~\ref{f08}{\sl  a}  shows  the ($\zeta,\zeta_{\rm  c}$)-plane  for
$\gamma_{\rm  c}=1.0$, $\eta=0.888$  and  without turbulent  pressure.
Although the  current analysis relies on simple  one-zone models, this
is a  very interesting  finding and we  are tempted to  attribute this
region of the plane $(\zeta,  \zeta_{\rm c})$ to the instability strip
of LPVs --- variable red  giants and supergiants --- which are thought
to  be significantly  nonadiabatic  and highly  convective.  There  is
general  agreement   within  the  astrophysical   community  that  the
intrinsic  reason for the  existence of  the red  edge of  the Cepheid
instability  strip is  the damping  produced by  convection.  However,
theoretical and empirical evidence  suggest that convection might also
be  a destabilizing mechanism  in the  cool region  of the  HR diagram
(\citealt{FW82,EG96,XDC98a,Wood00,CDKM01,DEA01,KB03,IEA04}).  When the
convective time  scale is  much longer than  the dynamical  time scale
($\zeta_{\rm c} \ll 1$), the  effect of convection is stabilizing, but
when the  convective time scale is  much shorter, the  reverse is true
(\citealt{G67}).  Thus, it may  provide an important driving mechanism
in  red   giants  and  supergiants.   More   exactly,  the  convective
luminosity  in the ionization  regions of  red pulsating  variables is
expected  to exceed  99\% of  the  total luminosity,  as mentioned  by
\cite{XDC98a}.   They identify  a Mira  instability strip  outside the
Cepheid  instability strip  when  pulsation-convection interaction  is
taken into account.

To investigate the  effect of the turbulent pressure in  the case of a
completely   convective    shell,   we   have    identified   in   the
($\zeta,\zeta_{\rm c}$)-plane the regions characterized by limit-cycle
behavior.  We  illustrate the  results in Fig.~\ref{f08}{\sl  b}.  The
region  of limit cycles  for the  completely convective  case persists
when the turbulent pressure is introduced.  Compared to the cases with
no turbulent  pressure, this  region shifts to  smaller values  of the
convective efficiency, $\zeta_{\rm c}$.  This shift is due to the fact
that  the  minimal  perturbation   strength  necessary  to  drive  the
pulsational  instability  is  achieved  at weaker  convective  driving
(smaller $\zeta_{\rm c}$), as the  rest of the driving is now provided
by the turbulent pressure.

\begin{figure}[t]
\epsfig{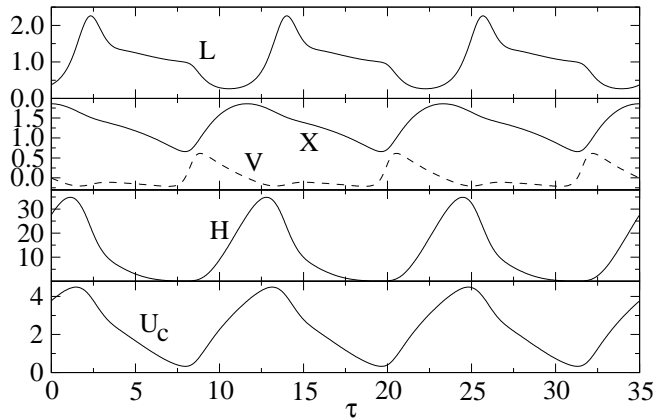}
\caption{Variation  of  the  luminosity ($L$),  radius  ($X$),
velocity ($V$), pressure ($H$), convective velocity ($U_{\rm c}$), for
the   case  of  completely   convective  shell   close  to   the  {\bf
vibrationally  unstable} edge  ($\zeta_{\rm c}=8.7$;  $\zeta=9$).  The
case of zero-turbulent pressure was considered. \label{f09}}
\end{figure}

Before  verifying  that  convection  is  the  driving  agent  for  the
completely  convective  shell,   we  briefly  describe  the  pulsation
characteristics  of the  models constructed  by  adopting $\gamma_{\rm
c}=1.0$ and  compare them with the previously  discussed limit cycles.
For comparative purposes, Fig.~\ref{f08}{\sl  c} shows the time series
of  a  ``Cepheid'' ($\gamma_{\rm  c}<0.5$),  and  of  a ``LPV''  model
($\gamma_{\rm  c}=1.0$). The  increase  both in  period and  amplitude
observed  in this  figure when  passing from  a ``Cepheid''-like  to a
``LPV''-like supports the previous working hypothesis. In this context
it  is  worth  mentioning  that ``Cepheid''-like  pulsators  show  the
well-known phase-lag  between the light  and the velocity  curve.  The
pulsation behavior of the ``LPV''-like  model needs to be discussed in
more detail. The data plotted in  the top and in the bottom panel show
a very small phase lag  between the light and the temperature changes.
This evidence  appears to be supported by  infrared spectroscopic data
(\citealt{HHR82,HSH84})  of pulsating  Asymptotic  Giant Branch  (AGB)
stars.  However, the maximum in the light curve takes place before the
minimum  in  the  velocity  curve.   This  finding  is  at  odds  with
observational  data  for Mira  and  Semiregular  Variables, since  the
minimum    velocity   is    not   correlated    with    light   maxima
(\citealt{HHR82,LKH00,LH02}).   Nevertheless,  the comparison  between
theory  and  observations  is  hampered  by  the  occurrence  of  long
secondary  periods  (\citealt{HLJF02,WOK04}),  and  indeed  in  a  few
objects the minimum  in the velocity curve takes  place later than the
maximum in the velocity  curve --- $\chi$~Oph, R~Leo (\citealt{HSH84})
and S~Lep  (\citealt{WOK04}).  In passing  we would also like  to draw
the attention  on the correlation between light  maxima and convective
velocity, since the maximum takes  place just before maximum light. We
are not aware of macroturbulent velocity measurements in LPVs, and new
observations would be very useful to constrain the plausibility of the
physical assumptions adopted in simple one-zone convective models.

For completeness, we illustrate in Fig.~\ref{f09} the time behavior of
a  model  located close  to  the  vibrational  instability edge  after
approaching to limit-cycle stability.  This example is generic for all
the  cases  with  or  without  the turbulent  pressure  close  to  the
vibrational-instability edge for  the completely convective shell. The
data  plotted  in this  figure  show  a  well-defined bump  along  the
decreasing  branch  of  the  light  curve.   This  bump  becomes  more
pronounced as the values of the parameters $\zeta$ and $\zeta_{\rm c}$
approach  the  vibrational-instability edge,  while  the  peak of  the
luminosity  becomes  a pulse-like  feature.   The  occurrence of  such
secondary features  is quite typical  along the light curves  of Miras
(\citealt{WOK04}).  Moreover, the occurrence of the luminosity maximum
appears  strongly  correlated with  the  variation  of the  convective
velocity ($U_{\rm c}$), while the bump  is in phase with the change of
the radial velocity ($V$).

Similar to the limit cycles for the weakly convective cases associated
to  the Classical  Cepheids,  it is  highly  speculative to  attribute
specific  values of  the  parameters $\zeta$  and  $\zeta_{\rm c}$  to
different  types  of  LPVs.   A  more detailed  investigation  of  the
parameter space  is mandatory to pinpoint the  pulsation properties of
completely convective one-zone models  that mimic an ``RGB''-like or a
``Mira''-like behavior.


\subsection{The $\gamma$-mechanism}

\begin{figure}[b]
\epsfig{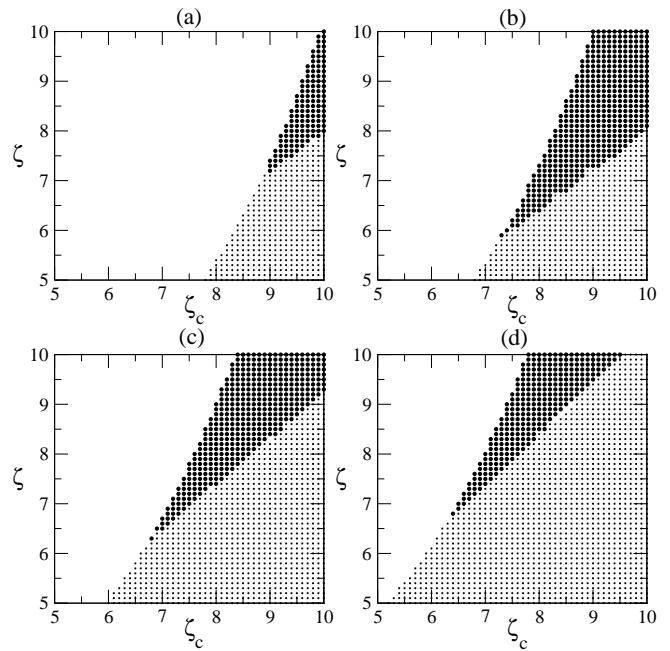}
\caption{The   ($\zeta,\zeta_{\rm   c}$)-plane   for  the   completely
convective shell ($\gamma_{\rm c}=1.0$) and different values of $\eta$
and $\Gamma_1$  giving birth to  several behaviors: limit  cycle ({\sl
filled  circles}),   pulsational  stability  ({\sl   dots})  and  {\bf
vibrational   instability}   ({\sl   white   regions}).    {\sl   (a)}
$\eta=0.92$,  $\Gamma_1=1.1$; {\sl (b)}  $\eta=0.888$, $\Gamma_1=1.1$;
{\sl  (c)}  $\eta=0.888$,   $\Gamma_1=1.2$;  {\sl  (d)}  $\eta=0.888$,
$\Gamma_1=1.3$. Eq.~(\ref{eq:dsystem}) has been used. \label{f10}}
\end{figure}

As  a  natural  further  step   into  clarifying  the  origin  of  the
pulsationally unstable  cases for the completely  convective shell, we
have investigated  the role of the adiabatic  exponent, $\Gamma_1$, as
driving  agent. From the  mathematical perspective  of the  model, the
question resides  into establishing the relative effect  on the system
dynamics of  the parameter $\gamma_{\rm c}$,  on one hand,  and of the
parameter $\Gamma_1$, on the other hand, included in the definition of
the    parameters    $q   \equiv    m    \Gamma_1-2$   and    $d\equiv
m(\Gamma_1-2)/2$. The  case of completely  convective shell translates
for   this   one-zone   model    into   the   disappearance   of   the
$\kappa$-mechanism,  with  the  driving  being supplied  only  by  the
$\gamma$-mechanism.   To test whether  the pulsational  instability of
our  convective  models is  due  to  the  $\gamma$-mechanism or  to  a
convection-induced driving mechanism, we  studied the evolution of the
limit-cycle region in  the ($\zeta,\zeta_{\rm c}$)-plane as $\Gamma_1$
is  varied.   The  results  obtained   so  far  have  used  the  value
$\Gamma_1=1.1$ as the one  employed by \cite{S86}.  We conjecture that
if the obtained cases of  self-sustained oscillations are a product of
convection-induced  driving,  then  the  $\gamma$-mechanism  has  very
little, if any, influence  on their properties.  Fig.~\ref{f10}{\sl a}
and {\sl b} display two cases of different shell thickness concluding,
as  in  $\S3.1$,  that  a  thicker  shell implies  a  wider  range  of
parameters  leading  to  pulsational  instability. In  the  completely
convective  case, this  increase  causes a  shift  of the  limit-cycle
region to lower $\zeta_{\rm c}$.   The same shift occurs by increasing
the $\Gamma_1$ value at constant  shell thickness, or, in other words,
by    decreasing   the    efficiency    of   the    $\gamma$-mechanism
(Fig.~\ref{f10}{\sl b,c,d}).  This means that convection and turbulent
pressure are driving mechanisms for fully convective models, since the
region where  these models display a limit  cycle stability marginally
depends  on the adopted  $\Gamma_1$ value.  This finding  supports the
theoretical  predictions  by   \cite{XDC98a}  concerning  the  driving
provided by the coupling between pulsation and convection, but we also
find that turbulent pressure is a destabilizing mechanism instead of a
damping factor for pulsation.  However, current models assume that the
physical  properties of  the  driving  region can  be  described by  a
Kramers' opacity law and by a very crude equation of state. Therefore,
there is  no guarantee  that current models  properly account  for the
thermodynamical  and  dynamical properties  of  the different  partial
ionization zones in low density envelope regions.


\section{Conclusions}

In  this paper  we have  thoroughly analyzed  the  one-zone convective
model introduced  by \cite{S86}.  The model  appears in the  form of a
system of  4 ordinary differential  equations where the  variables are
the radius of the shell, the velocity, the pressure and the convective
velocity.   The  nonadiabaticity  resides  in the  pressure  which  is
considered as  a nonadiabatic  perturbation of the  reference pressure
and, thus, can be considered as a nonadiabatic variable of the system.
The  model accounts for  the self-excited  oscillations by  adopting a
value  for the adiabatic  exponent, $\Gamma_1$,  close to  unity.  The
main parameters  of the  system are the  fraction $\gamma_{\rm  c}$ of
convective luminosity  with respect to  the total luminosity,  and the
time-scales ratios,  $\zeta$ and $\zeta_{\rm c}$, which  are a measure
of  the dynamical  time scale  to the  thermal time  scale and  to the
convective time  scale, respectively.  We  have extended the  model by
considering explicitly the role of the turbulent pressure and the case
of completely convective shell.

The one-zone models constructed by \cite{S86} were integrated only for
a  few   dynamical  time  scales  and,  therefore,   the  approach  to
limit-cycle stability remained still  to be investigated.  To properly
identify the region  of the parameter space which  shows a limit-cycle
behavior  (pulsational  instability), we  have  selected a  parametric
space given by $\gamma_{\rm c}  \in [0,1)$, $\zeta$ and $\zeta_{\rm c}
\in [0,10]$.  Our parametric study revealed well-defined regions where
limit cycles exist, born through  the Hopf bifurcation.  For a typical
shell thickness of 11\%,  pulsational instability was encountered only
for  the radiation-dominated cases  ($\gamma_{\rm c}<0.5$),  while the
increase in the  shell thickness displaces this limit  to upper values
of $\gamma_{\rm c}$.  The existence of these upper values supports the
work  by  \cite{S86}  and  further  strengthens  the  role  played  by
convection as a damping mechanism.   Moreover, we found that the shell
thickness  is,  as expected,  positively  correlated  with the  period
distribution of  pulsationally unstable cases.   Interestingly enough,
models constructed by adopting thick and thin shells obey different PL
relations that  mimic the behavior  of fundamental and  first overtone
Cepheids.

Additionally, we have undertaken a parametric study to investigate the
influence of  the turbulent  pressure on the  overall dynamics  of the
system and implicitly on the existence of limit cycles.  The turbulent
pressure appears  to be a driving  mechanism as the  regions where the
models  approach a limit-cycle  stability are  more extended  than for
models that do  not account for the turbulent  pressure.  Moreover and
even more importantly,  we find that one-zone models  that account for
turbulent pressure do not show a spurious peak along the rising branch
of light curves.  The inclusion of  this term is also supported by the
occurrence  in the  ($\zeta,\zeta_{\rm c}$)-plane  of  a pulsationally
unstable  region  bounded  to  the  left  and  to  the  right  by  two
pulsationally stable regions, as expected from physical considerations
and hydrodynamical models. Moreover, the models located in this region
display a morphology of the  light curves and the velocity time series
quite similar to those  of nonlinear, hydrodynamical models and actual
Cepheids.

As  a  natural  continuation  of  the  work  by  \cite{S86},  we  have
investigated the  vibrational and pulsational  stability of completely
convective models.  Our predictions  disclose a well-defined region of
the  parameter space  where these  red models  approach  a limit-cycle
stability. The turbulent pressure appears to be a driving mechanism, a
finding  that  supports  the   results  original  brought  forward  by
\cite{OC86} and \cite{CO93} on the  basis of both linear and nonlinear
LPV models, but  at odds with predictions based  on linear, convective
models provided  by \cite{XDC98a}.   We computed several  sequences of
models by adopting  different values of the adiabatic  exponent and of
the shell  thickness.  We found  in agreement with  \cite{XDC98a} that
the  coupling between  pulsation and  convection is  the  key physical
mechanism  that  drives  the  pulsation instability  in  these  simple
structures.   We have  also  performed a  qualitative comparison  with
empirical properties  of LPVs.  We  have found some  similarities, but
only one feature partially agrees  with empirical data: the maximum in
the light  curve is  not correlated with  the minimum in  the velocity
curve.  However, the comparison with observations might be hampered by
the  occurrence  of long  secondary  periods.   We  are not  aware  of
detailed measurements  of macroturbulent velocity  along the pulsation
cycle  of LPVs, however  our models  suggest that  the maximum  in the
light  curve takes  place soon  after  the maximum  in the  convective
velocity.

Current one-zone  models account  for the coupling  between convection
and  pulsation  and  for  turbulent  pressure.   We  investigated  the
sensitivity to free parameters and to the physical assumptions adopted
to construct the  models. However, the treatment and  the inclusion of
these physical  ingredients rely on crude  physical approximations.  A
more detailed  investigation is required before we  can assess whether
our current  theoretical framework  might mimic the  complex pulsation
behavior  of  LPVs  and  this   implies  the  use  of  the  asymptotic
perturbation  theory to  compute  analytically the  properties of  the
limit cycle.

The pulsational behavior  disclosed by  these simple  models for
fully convective  models will be  addressed in a forthcoming  paper on
the basis of nonlinear,  hydrodynamical models (Munteanu et al.  2005,
in preparation).

\acknowledgments It  is a  pleasure to thank  P.  Wood for  a detailed
reading  of a  draft of  this manuscript  and for  several suggestions
concerning the comparison between  theory and observations.  This work
has been  partially supported by the  MCYT grant AYA2002-04094-C03-01,
by the European Union FEDER funds, and by the CIRIT.  One of us, G.B.,
acknowledges  for partial  support INAF  within the  framework  of the
project:  ``The  Large  Magellanic  Cloud  a  laboratory  for  stellar
astrophysics``,  and PRIN~2003  within the  framework of  the project:
``Continuity      and       Discontinuity      in      the      Galaxy
Formation``. A.M. acknowledges a  scholarship of UPC which allowed her
to conclude the present work.

\begin{appendix}

\section{The turbulent pressure}

In order to  include the turbulent pressure and  its associated energy
terms in the present one-zone  model, one must start from the equation
of momentum and energy conservation for the treatment of convection of
\cite{S82a}:

\begin{eqnarray}
&&\frac{D\langle\mbox{\boldmath$u$}\rangle}{Dt}~=~-\frac{1}{\rho}~
\nabla(P+P_{\rm t})-\nabla\Phi\label{eq:momenteq}\\
&&\frac{D}{Dt}~(E+E_{\rm t})+(P+P_{\rm t})~\frac{DV}{Dt}~=~
-\frac{1}{\rho}\nabla\cdot(F_{\rm r}+F_{\rm c}+F_{\rm t}).
\label{eq:energy}
\end{eqnarray}

\noindent  Here as  well  as throughout  our  work, we  have used  the
notations employed in the original investigations. In \cite{S82a}, any
quantity was written  as $x=\langle x \rangle +  x'$, where $\langle x
\rangle$ was  the mean  quantity and $x'$,  the fluctuating  part.  In
Eq.~(\ref{eq:energy}),   $D/Dt=(\partial   /\partial   t   +   \langle
\mbox{\boldmath$u$}  \rangle  \cdot \nabla)$  is  the Lagrangian  time
derivative with $u$ being the convective velocity, $E$ is the specific
internal energy, $V\equiv  1/\rho$ is the specific volume,  $P$ is the
thermodynamic pressure and

\begin{eqnarray}
&& E_{\rm t}\equiv\frac{1}{2}\langle(u')^2\rangle,\\
&& P_{\rm t}\equiv\rho\langle(u')^2\rangle,\\
&& F_{\rm c}\equiv\rho C_{\rm p}\langle(u'T')\rangle,\\
&& F_{\rm t}\equiv\frac{1}{2}\langle(\mbox{\boldmath$u'$})^2 
\mbox{\boldmath$u$}
\rangle \label{eq:Ft}
\end{eqnarray}

\noindent   represent,  respectively,   the  convective   energy,  the
turbulent pressure,  the convective  and turbulent kinetic  fluxes, as
they are defined  in \cite{S82a}.  These equations must  be adapted to
the one-zone model.  The momentum equation translates into

\begin{equation}
\frac{d^2 X}{d\tau^2}=(1-\alpha_{\rm p})X^{-q}h+\alpha_{\rm p}X^{-c}
U_{\rm c}^2-X^{-2},
\label{eq:momentum}
\end{equation}

\noindent  where  Eq.~(\ref{eq:turbulence})  has  been  used  and  the
parameters are $q\equiv m\Gamma_1-2$, $c\equiv m-2$ and

\begin{equation}
\alpha_{\rm p}\equiv\frac{P_{{\rm t}_0}}{P_0+P_{{\rm t}_0}}
=\frac{X_0^{-m}U_{{\rm c}_0}^2}{X_0^{-m\Gamma_1} h_0+X_0^{-m} U_{{\rm
c}_0}^2}~. 
\label{eq:ap}
\end{equation}

\noindent To ease the calculation,  the energy equation can be divided
into

\begin{eqnarray}
\frac{\partial(L_{\rm r}+L_{\rm c})}{\partial m}&=& \frac{\Gamma_1
P}{\rho^2(\Gamma_3-1)}~\frac{\partial\rho}{\partial t}-
\frac{1}{\rho(\Gamma_3-1)}~\frac{\partial P}{\partial t}
\label{eq:LrLc} \\
\frac{\partial L_{\rm t}}{\partial m}&=& \frac{P_{\rm
t}}{\rho^2}~\frac{\partial\rho}{\partial t} -\frac{\partial E_{\rm
t}}{\partial t} .
\label{eq:Lt}
\end{eqnarray} 

\noindent  In  the  determination  of  the  convective  and  turbulent
luminosities, one  can use the  conservative choice of $L_{\rm  c}$ of
Eq.~(27) of  \cite{S86} and Eq.~(\ref{eq:Ft}).   The calculations lead
to an  expression for  $L_{\rm t}$ identical  to that of  $L_{\rm c}$,
that  is  $L_{\rm  t}=X^{-c}  U_{\rm c}^3$,  while  $E_{\rm  t}=U_{\rm
c}^2/2$.  Using  Eqs.~(\ref{eq:LrLc}--\ref{eq:Lt}),  one can  get  the
final  equation  of  energy   conservation  taking  into  account  the
turbulence:

\begin{eqnarray}
\frac{dh}{d\tau}=&-&m~\frac{\alpha_{\rm p}(\Gamma_3-1)}{1-\alpha_{\rm
p}}X^{2d-1}~\frac{dX}{d\tau}~U_{\rm c}^2~-~\frac{\rho_0(\Gamma_3-1)}
{P_0}~\zeta_{\rm c}~(X^d h^{1/2}U_{\rm c}-\nonumber\\     
&-&X^{2d}U_{\rm c}^2)~-~\zeta X^{2d}~(\gamma_{\rm r} X^b h^{s+4}
+(1-\gamma_{\rm  r}) X^{-c} U_{\rm c}^3 -1) ,
\label{eq:energyeq}
\end{eqnarray}

\noindent where  $d \equiv m (\Gamma_1  -1) / 2$ and  $ \gamma_{\rm r}
\equiv  L_{{\rm  r}_0}  /  L_0$.    All  the  other  symbols  used  in
Eq.~(\ref{eq:energyeq})  have their usual  meaning.  The  equation for
the convective velocity coincides with the zero-turbulence case as the
approximation  made for  its recovery  uses  of the  temperature as  a
function of the thermodynamic pressure  only. In the case in which the
turbulent energy, $E_{\rm  t}$, is neglected in Eq.~(\ref{eq:energy}),
Eq.~(\ref{eq:energyeq}) becomes

\begin{eqnarray}
\frac{dh}{d\tau}=&-&m~\frac{\alpha_{\rm p}(\Gamma_3-1)}{1-\alpha_{\rm
p}}~X^{2d-1}~\frac{dX}{d\tau}~U_{\rm c}^2~-\nonumber\\
&-&~\zeta X^{2d}~(\gamma_{\rm r}X^b h^{s+4}+(1-\gamma_{\rm  r})X^{-c}
U_{\rm c}^3 -1).
\label{eq:energysimple}
\end{eqnarray}

Thus,  to a first  approximation, the  one-zone convective  model with
turbulent  pressure  is   described  by  Eqs.~(\ref{eq:momentum})  and
(\ref{eq:energysimple}),    while   for   the    convective   velocity
Eq.~(\ref{eq:dsystem}) remains valid.

\end{appendix}

\end{document}